\documentclass[10.5pt]{article}
\usepackage{amsfonts}

\def\sqr#1#2{{\vcenter{\vbox{\hrule height.#2pt
              \hbox{\vrule width.#2pt height#1pt \kern#1pt \vrule width.#2pt}
              \hrule height.#2pt}}}}
\def\3n{\negthinspace \negthinspace \negthinspace }
\def\2n{\negthinspace \negthinspace }
\def\1n{\negthinspace }

\def\={\buildrel \triangle \over =}

\def\ms{\medskip}

\def\({\Big (}
\def\){\Big )}
\def\[{\Big[}
\def\]{\Big]}
\def\bde{\begin{definition}}
\def\ede{\end{definition}}
\def\be{\begin{equation}}
\def\bel{\begin{equation}\label}
\def\ee{\end{equation}}
\def\bt{\begin{theorem}}
\def\et{\end{theorem}}
\def\bc{\begin{corollary}}
\def\ec{\end{corollary}}
\def\bl{\begin{lemma}}
\def\el{\end{lemma}}
\def\bp{\begin{proposition}}
\def\ep{\end{proposition}}
\def\bas{\begin{assumption}}
\def\eas{\end{assumption}}
\def\br{\begin{remark}}
\def\er{\end{remark}}
\def\ba{\begin{array}}
\def\ea{\end{array}}
\def\ed{\end{document}}

\def\square#1{\vbox{\hrule\hbox{\vrule height#1%
     \kern#1\vrule}\hrule}}
\def\rectangle#1#2{\vbox{\hrule\hbox{\vrule height#1%
     \kern#2\vrule}\hrule}}

\font\tenbb=msbm10 \font\sevenbb=msbm7 \font\fivebb=msbm5

\newfam\bbfam
\scriptscriptfont\bbfam=\fivebb \textfont\bbfam=\tenbb
\scriptfont\bbfam=\sevenbb

\newtheorem{lemma}{Lemma}[section]
\newtheorem{remark}{Remark}[section]
\newtheorem{example}{Example}[section]
\newtheorem{theorem}{Theorem}[section]
\newtheorem{corollary}{Corollary}[section]

\newtheorem{definition}{Definition}[section]
\newtheorem{proposition}{Proposition}[section]
\newtheorem{assumption}{Assumption}[section]

\makeatletter
   
   \@addtoreset{equation}{section}
\makeatother

\usepackage{amsfonts}
\usepackage{latexsym}
\usepackage{amsmath}
\usepackage{amssymb}
\usepackage{color}

\begin{document}

\title{Risk minimizing of derivatives via dynamic $g$-expectation and related
topics\footnote{This work is supported in part by National Natural
Science Foundation of China (Grants 10771122 and 11071145), Natural
Science Foundation of Shandong Province of China (Grant Y2006A08),
Foundation for Innovative Research Groups of National Natural
Science Foundation of China (Grant 10921101), National Basic
Research Program of China (973 Program, No. 2007CB814900),
Independent Innovation Foundation of Shandong University (Grant
2010JQ010), Graduate Independent Innovation Foundation of Shandong
University (GIIFSDU) and the 111 Project No. B12023.  }}

\author{Tianxiao Wang \footnote{School of Mathematics, Shandong
University, Jinan 250100, China. Part of the work was done while the
author was visiting Department of Mathematics, University of Central
Florida, USA. The author would like to thank Professor Jiongmin Yong
for his hospitality.}}

\maketitle

\begin{abstract}
In this paper, we investigate risk minimization problem of
derivatives based on non-tradable underlyings by means of dynamic
$g$-expectations which are slight different from conditional
$g$-expectations. In this framework, inspired by \cite{AID2010} and
\cite{OS2009MF}, we introduce risk indifference price, marginal risk
price and derivative hedge and obtain their corresponding explicit
expressions. The interesting thing is that their expressions have
nothing to do with nonlinear generator $g$, and one deep reason for
this is due to the completeness of financial market. By giving three
useful special risk minimization problems, we obtain the explicit
optimal strategies with initial wealth involved, demonstrate some
qualitative analysis among optimal strategies, risk aversion
parameter and market price of risk, together with some economic
interpretations.

\end{abstract}

\ms


\ms


\ms

\section{Introduction}

Recently there are many financial instruments written on
non-tradable underlyings such as weather future, catastrphe future,
and other financial products. Since they are impossible to perfectly
hedge, one has to look for some well correlated tradable assets to
cross hedge the risk. As to the pricing and hedging problem for such
derivatives, see for example Ankirchner et al (\cite{AID2010},
\cite{AIP2008}). In this paper, we propose a new framework within
which to address the dynamic risk minimization problem of above
derivatives.

As to the risk minimization problem, in the literature there are
various form of works on it, such as Mataramvura and ${\O}$ksendal
\cite{MO2008} in a zero-sum stochastic differential game framework,
${\O}$ksendal and Sulem \cite{OS2009} via stochastic maximum
principle for FBSDEs, Horst and Moreno-Bromberg \cite{HM2008} in a
principle agent game framework. In this paper, we consider this
problem by a new important representation of dynamic convex risk
measures via BSDEs. As we know, $g$-expectations and conditional
$g$-expectations were introduced by Peng \cite{P1997} as solutions
of a class of nonlinear BSDEs and proved by Rosazza Gianin
\cite{R2006} to provide examples of dynamic coherent or convex risk
measures under suitable hypothesis. Therefore, inspired by this, we
will use a slight different way to represent dynamic risk measures.
More precisely, given $t\in[0,T]$, $x$ being initial wealth at time
$t$, $X^{t,x}(T)$ being the terminal wealth, if the following BSDE
admits a unique solution $(Y,Z)$ on $[t,T]$,
\begin{equation}\label{0}
Y(r)=-X^{t,x}(T)+\int_r^Tg(s,Z(s))ds-\int_r^TZ(s)dW(s),\quad
r\in[t,T],
\end{equation}
we then define $\rho(t;X^{t,x}(T))=Y(t)$ with $t\in[0,T]$ the
dynamic $g$-expectation on $[0,T]$, $\rho_1(r;X^{t,x}(T))=Y(r)$ with
$r\in[t,T]$ the conditional $g$-expectation on $[t,T].$
Particularly, if $t=0$, the above two notions degenerate into the
classical $g$-expectations (denoted by $\rho_2(0;X^{0,x}(T))$) and
conditional $g$-expectations (denoted by $\rho_3(r;X^{0,x}(T))$ with
$r\in[0,T]$) in \cite{P1997} and \cite{R2006}. Note that $\rho$,
$\rho_1$ and $\rho_3$ above can be regarded as dynamic risk measures
via BSDE (\ref{0}).

In this paper, we will make use of dynamic $g$-expectation $\rho$,
rather than $\rho_1$ and $\rho_3$, to study the risk minimization
problem (see Section 2 below). This is not just because of the
natural form of $\rho$, while some deeper meanings are involved.
Actually, in order to make the results meaningful from the economic
point of view, Bj\"{o}rk et al \cite{BMZ2012} argued that risk
aversion parameter (or risk tolerance parameter) should be dependent
on initial wealth in the dynamic mean-variance portfolio
optimization, see also Hu et al \cite{HJZ2012}, which inspires us to
consider the risk measures in more general yet realistic framework.
For example, we can represent dynamic convex risk measure by dynamic
$g$-expectation above in the form
\begin{eqnarray}\label{1}
\rho(t,X^{t,x}(T))=-X^{t,x}(T)+\int_t^Tg(s,x,Z(s))ds-\int_t^TZ(s)dW(s),
\end{eqnarray}
with $g$ being convex and depending on initial wealth $x$. In
particular, when $g(s,x,z)=\frac{1}{2\delta(x)}|z|^2$,
$$\rho(t;X^{t,x}(T))=\delta(x)\log
\mathbb{E}^{\mathcal{F}_t}\exp\left(-\frac{X^{t,x}(T)}{\delta(x)}\right)
,\quad t\in[0,T],\quad x\in \mathbb{R},$$with $\delta(x)$ being the
risk tolerance parameter, is a generalized dynamic entropic risk
measure. Similarly ideas also appear in Example 3.1 below. On the
other hand, if we similarly define dynamic risk measure $\rho_1$
above on $[t,T]$ in such setting, it will become controversial with
one basic axiom of defining dynamic risk measure, that is,
independence of the past, since $g$ depends on initial wealth $x$.
For example, if $t=0,$ and risk aversion parameter $\gamma$ is a
decreasing function in $x$, then it becomes unsuitable to use
conditional $g$-expectation in $[0,T]$ to represent dynamic risk
measures.

In this paper we mainly consider the case of $g$ being Lipschitz in
$z$. If $X^{t,x}(T)$ is the terminal wealth depending on some
control variable or investment strategy (see for example (\ref{2})
below), our aim is then to find an optimal strategy to minimize the
risk defined in (\ref{1}) in a dynamical manner. For this risk
minimization problem, we firstly give one sufficient condition for
the existence of optimal strategy by means of comparison theorem for
BSDEs. From a simple example, we see that a new minimizing problem,
playing a key role in above sufficient condition, is also necessary
in some sense. This inspires us to study this minimizing problem
more deeply. By using Legendre-Fenchel transform and
sub-differential for convex functions, we give one sufficient and
necessary condition for this minimization problem, and explain it by
two simple examples. Furthermore, we use the procedure here to
derive risk indifference price, marginal risk price and derivative
hedge, the definitions of which are inspired by the work in
\cite{AID2010} and \cite{OS2009MF}. At last by giving three special
and useful cases of $g$ in (\ref{1}), we study the corresponding
risk minimization problem and get the explicit form of optimal
strategy. As in \cite{BMZ2012}, the strategy here depends on initial
wealth and their economic rationality via qualitative analysis are
also emphasized. We talk about the tight relations between the
existence of optimal solution and two basic financial notions, that
are, risk aversion parameter and market price of risk.

The main novelty of this paper contains the following: Firstly we
introduce two explicit kinds of risk measures via dynamic
$g$-expectation representing two classes of investors with different
attitude to risk. By looking for some subjective parameter of
investor and studying its connection with market price of risk
(objective parameter), we explore the requirements of wellposedness
of optimal strategy. Some comparisons between these two cases are
also given with some reasonable interpretation. Secondly, inspired
by the recent work of \cite{S2009} and \cite{S2010}, when risk
measures are expressed by BSDEs with generators being regarded as
continuous time analogue of discrete Gini principle, we derive the
explicit optimal strategy for the corresponding risk minimization
problem, identify the term representing risk aversion parameter.
Note that the form of $\gamma(x)\cdot g$ here, with $\gamma(x)$
being the risk aversion parameter, is just the well-known Huber
penalty function in \cite{H1964}. To our best knowledge, it is the
first time to do analysis in above two aspects via dynamic
$g$-expectation. Thirdly, when dealing with above special risk
minimization problems, we also give some qualitative analysis among
optimal strategy, market price of risk, initial wealth, and risk
aversion parameter. In other words, we talk about the connections
between market price of risk and optimal strategy, compare optimal
strategy across different investors and wealth level, which are
consistent with the results in \cite{B2007}, \cite{PZ2012} and
\cite{X2011}. Fourthly, when $g$ is Lipschitz in $z$, we not only
derive the explicit form of risk indifference price, marginal risk
price and derivative hedge, but also notice one interesting
phenomenon, that is their independence of $g$ representing risk
preference of investors. By comparing with \cite{AID2010}, we
believe that one potential reason for this is the completeness of
financial market. Such result obtained via BSDEs is still new to our
knowledge.

The paper is organized as follows: In Section 2, we will formulate
the risk minimization problem and give some useful notations. In
Section 3, we will give one sufficient condition to ensure the
existence of optimal strategy. By exploring a minimizing problem
especially, we derive the explicit form of risk indifference price,
marginal risk price and derivative hedge. Some examples are given to
further explain these results. In Section 4, we study three
important and inspiring cases by connecting with risk aversion
parameter and market price of risk, exploring some qualitative
analysis for the expressions and explaining the economic rationality
of obtained results.

\section{Model formulation and preliminary}
Throughout this paper, we let
$(\Omega,\mathcal{F},\mathbb{F},\mathbb{P})$ be a complete filtered
probability space on which a one-dimensional standard Brownian
motion $W(\cdot)$ is defined with
$\mathbb{F}=\{\mathcal{F}_t\}_{t\geq 0}$ being its natural
filtration augmented by all the $\mathbb{P}$-null sets. Suppose that
the price of a one-dimensional non-tradable index (such as a stock,
temperature or loss index) follows the dynamic
$$
dR^{t,r_0}_s=b(s,R^{t,r_0}_s)dt+\sigma (s,R^{t,r_0}_s)dW_s,\quad
s\in[t,T],\ r_0\in\mathbb{R},
$$
with $t\in[0,T],$ $b:[t,T]\times \mathbb{R}\rightarrow \mathbb{R}$
and $\sigma :[t,T]\times \mathbb{R}\rightarrow \mathbb{R}$ being
Lipschitz and linear growth. For simplicity of presentation, we
assume that in the correlated financial market there are one
non-risky asset used as numeraire, and one risky asset, the price of
which evolves according to
$$
dS(r)=S(r)[\alpha (r,R_r^{t,r_0})dr+\beta (r,R_r^{t,r_0})dW(r)],
$$
where $\alpha $ is bounded and $\beta $ satisfies $\epsilon \leq
|\beta |\leq K$ with $0<\epsilon <K$ being two constants. Provided
the investor
invests $\pi _t$ in risky asset at time $t,$ the wealth of investor at time $%
s$ conditional on the wealth $x$ at time $t$ and the index $R_t=r_0$
is given by
\begin{eqnarray}\label{2}
X^{\pi ,t,r}(s)=x+\int_t^s\pi (u)[\alpha (u,R^{t,r_0}_u)du+\beta
(u,R^{t,r_0}_u)dW(u)].
\end{eqnarray}
In the following without special explanation we simplify $X^{\pi
,t,r}(\cdot)$ by $X^{\pi}(\cdot)$. The investment strategy $\pi $ is
called admissible if $\mathbb{E}\int_t^T|\pi (s)\sigma
(s,R^{t,r_0}_s)|^2ds<\infty \ $and we denote the set of admissible
strategies by $\Pi^{t,r_0}.$

The above model is adapted from the one in \cite{AID2010}, where
pricing and hedging principles for derivative based on nontradable
underlyings are discussed in the framework of exponential utility
maximization. In this paper we will study this model from a new view
of minimizing the risk of derivatives by using dynamic $g$%
-expectation mentioned in the introduction. More precisely, given
$t\in[0,T],$ $X^{\pi}(T)$ of (\ref{2}), and $(\rho (t,X^\pi
(T)),Z^{\pi}(\cdot))$ satisfying
$$
\rho (t,X^\pi (T))=-X^\pi
(T)+\int_t^Tg(s,x,Z(s))ds-\int_t^TZ(s)dW(s)
$$
with $x$ being the initial wealth of $X^{\pi}(T)$, our risk
minimization problem can be posed as follows.
\begin{eqnarray*}
&& \text{\rm \bf Problem(RM).}\ \ \text{\rm Find a }\
\overline{\pi}\in
\Pi^{t,r_0}\ \text{\rm such that} \\
&&\quad \quad \quad \rho _t(X^{\overline{\pi }}(T))=\rho
(t,X^{\overline{\pi }}(T))=\mathop{\rm essinf}_\pi \rho (t,X^\pi
(T)),\text{}t\in [0,T].
\end{eqnarray*}
Since the generator depends on $x$; we are looking for optimal
strategy of the form $\pi \doteq \pi (t;x).$ Besides the optimal
strategy, we will introduce three related financial notions. Given a
bounded and measurable function $F:\mathbb{R}\rightarrow
\mathbb{R}$, suppose an investor receives a derivative of form
$F(R_T^{t,r_0})$ kept in his portfolio until maturity $T$ where the
corresponding optimal strategy, denoted by $\widehat{\pi }(t;x)$
satisfies
\[
\rho _t(X^{\widehat{\pi }}(T)+F(R_T^{t,r_0}))=\mathop{\rm
essinf}_\pi \rho (t;X^\pi (T)+F(R_T^{t,r_0}));\quad t\in [0;T].
\]
Obviously the appearance of derivative $F(R_T^{t,r_0})$ leads to a
change in
the optimal strategy. We then call the difference $\Delta =\widehat{\pi }-%
\overline{\pi }$ derivative hedge, which is required to hedge the
risk associated with the derivative in the portfolio.

Suppose the investor spends a payment $q\doteq q(t,x)$ for derivative $%
F(R_T^{t,r_0})$, we thus give the notion of risk indifferent price.
\begin{definition}
 We call $q$ the dynamic risk indifference price of the
derivative $F(R_T^{t,r_0})$ at time $t$, if it is the solution of
the equation
\begin{eqnarray*}
\mathop{\rm essinf}_\pi \rho _t(X_{x-q}^\pi (T)+F(R_T^{t,r_0})) &=&\rho _t(X_{x-q}^{%
\widehat{\pi }}(T)+F(R_T^{t,r_0})) \\
&=&\rho _t(X_x^{\overline{\pi }}(T))=\mathop{\rm essinf}_\pi \rho
_t(X^\pi_x(T));
\end{eqnarray*}
\end{definition}
One important notion related to risk indifference price is marginal
risk price while the difference price is nonlinear in the sense that
risk indifference price of $k\cdot F(R^{t,r_0}_T)$ does not equal to
$k$ (natural number) times the indifference price of
$F(R^{t,r_0}_T)$. We denote marginal risk price by $p(t,x)$. After
paying $p(t,x)$ for the derivative, the investor is indifferent
between buying or not buying an infinitesimal amount of derivative.
Similar ideas also appears in Section 6 in \cite{AID2010} where the
notion of marginal utility price is introduced.

\section{Risk minimization problem and related topics}

In this section, we will study {\bf Problem (RM)} and three related
financial notions mentioned above. By making use of comparison
theorem of BSDEs we will derive one sufficient condition for risk
minimization problem, while similar ideas appeared in \cite{HPD2010}
for the special case of $t=0$ and $g$ being independent of $x$. We
denote by $\mathcal{P}^{t,r}$ the set of process
$$p(\cdot ):=p(\cdot ,R^{t,r_0}(\cdot ),x)=\pi (\cdot ,x)\beta (\cdot
,R^{t,r_0}(\cdot ))$$ with $\pi (\cdot ,x)\in \Pi^{t,r_0}.$ For
given $t\in[0,T]$, $r\in[t,T]$, the wealth process in (\ref{2}) can
be rewritten as
\begin{eqnarray}\label{2.1}
X^{p}(r)=x+\int_t^r\theta
(u,R_u^{t,r_0})p(u,R_u^{t,r_0},x)du+\int_t^rp(u,R_u^{t,r_0},x)dW(u).
\end{eqnarray}
For reason of simplicity, we consider the special case of $F=0.$
Given $t\in[0,T]$, from Section 1, the corresponding risk
minimization problem is to minimize $\rho(t;X^{p}(T))$, where
\begin{eqnarray}\label{2.2}
\rho(t;X^{p}(T))=-X^p(T)+\int_t^Tg(s,x,Z^{p,t}(s))ds-\int_t^TZ^{p,t}(s)dW(s).
\end{eqnarray}
After observing the fact of
\begin{eqnarray*}
&&-X^p(T)=-X^p(r)-\int_r^Tp(s,R_s^{t,r_0},x)\theta
(s,R_s^{t,r_0})ds-\int_r^Tp(s,R_s^{t,r_0},x)dW(s),\\
 && Y^{p,t}(r)
=-X^p(r)+\int_r^T[g(s,x,Z^{p,t}(s))-p(s,R_s^{t,r_0},x)\theta
(s,R_s^{t,r_0})]ds \\
&&\ \ \ -\int_r^T[Z^{p,t}(s)+p(s,R_s^{t,r_0},x)]dW(s),
\end{eqnarray*}
we arrive at
\[
\overline{Y}^{p,t}(r)=-x+\int_r^T\overline{g}(s,x,\overline{Z}%
^{p,t}(s),p(s,R_s^{t,r_0},x))ds-\int_r^T\overline{Z}^{p,t}(s)dW(s),\text{ }%
r\in [t,T],
\]
where
\begin{eqnarray*}
\overline{Y}^{p,t}(r) &=&Y^{p,t}(r)+\int_t^rp(u,R_u^{t,r_0},x)\theta
(u,R_u^{t,r_0})du, \\
&&\ +\int_t^rp(u,R_u^{t,r_0},x)dW(u), \\
\overline{Z}^{p,t}(s) &=&Z^{p,t}(s)+p(s,R_s^{t,r_0},x), \\
\overline{g}(s,x,\overline{Z}^{p,t}(s),p(s,R_s^{t,r_0},x)) &=&g(s,x,\overline{Z%
}^{p,t}(s)-p(s,R_s^{t,r_0},x))-p(s,R_s^{t,r_0},x)\theta
(s,R_s^{t,r_0}).
\end{eqnarray*}
It follows from the comparison theorem for BSDEs that
\begin{theorem}
For given $t$, $s,$ $x,$ $z,$ suppose that there is one
$\overline{p}\doteq \overline{p}(\cdot,R^{t,r_0}(\cdot),x)\in
\mathcal{P}^{t,r_0}$ such that
\begin{eqnarray}\label{3}
\overline{g}(s,x,z,\overline{p}(s,R_s^{t,r_0},x))=\mathop{\rm essinf}_p\overline{g}%
(s,x,z,p(s,R_s^{t,r_0},x)),\text{ }s\in [t,T],\text{ }x\in \Bbb{R.}
\end{eqnarray}
Moreover, $g$ is Lipshitz in $z.$ Then
\[
\overline{Y}^{\overline{p},t}(r)=\mathop{\rm essinf}_p\overline{Y}^{p,t}(r)\text{ with }%
r\in [t,T].
\]
In particular, $\overline{Y}^{\overline{p},t}(t)=\mathop{\rm essinf}_p\overline{Y}%
^{p,t}(t).$ Furthermore, if $g$ is convex and differentiable in $z$
such that
\[
\overline{g}_p(s,x,\overline{Z}^{\overline{p},t}(s),\overline{p}%
(s,R_s^{t,r_0},x))=0,\text{ }s\in [t,T],
\]
then $$\overline{\pi }(s,x)=\overline{p}(s,R_s^{t,r_0},x)\beta
^{-1}(s,R_s^{t,r_0}),\quad s\in[t,T],\ x\in R,$$ is an optimal
strategy of {\bf Problem(RM)}.
\end{theorem}
In Section 4 we will give three important examples to illustrate the
useful application of Theorem 3.1. Before that let us look at
another simple yet interesting example, from which we can see the
necessity of condition (\ref{3}) above. Here for $s\in [t,T],$ $x\in
\Bbb{R}$, $\omega \in \Omega ,$ we assume that
$p(s,R_s^{t,r_0},x)\in \Gamma \doteq \Gamma _s(\omega )$ with
$\Gamma _s(\omega )\subseteq \Bbb{R}$ being closed and convex. For
given $a\in\mathbb{R},$ we denote by $\mathop{\rm dist}_{\Gamma}(a)$
the distance between $a$ and $\Gamma$, and $\Psi_{\Gamma}(a)$ the
element of $\Gamma$ such that
$\left|a-\Psi_{\Gamma}(a)\right|=\mathop{\rm dist}_{\Gamma}(a).$
\begin{example}
\rm
 Consider BSDE of
\[
Y^{p,t}(r)=-X^{p,t}(T)+\int_r^T\frac{\gamma (x)}{2}|Z^{p,t}(s)|^2ds-%
\int_r^TZ^{p,t}(s)dW(s),r\in [t,T],
\]
where $X^{p,t}(T)$ satisfies (\ref{2.1}), $\gamma $ is a positive
and decreasing function of initial wealth $x$ representing risk
aversion parameter. Note that $Y^{p,t}(\cdot)$ is also a dynamic
exponential utility function with $\gamma(x)$, that is,
\[
Y^{p,t}(r)=\frac{1}{\gamma (x)}\ln \mathbb{E}^{\mathcal{F}_r}\left[
\exp \left( -\gamma (x)
X^{p,t}(T)\right) \right] ,\text{ }r\in [t,T],\text{ }x\in \Bbb{%
R}\text{.}
\]
Our problem is to find an optimal strategy $\overline{\pi}$ to
minimize $Y^{p,t}(t)$ with $t\in[0,T].$ In this case, we have
\begin{eqnarray}\label{4}
&&\ \ \overline{g}(s,x,\overline{Z}^{p,t}(s),p(s,R_s^{t,r_0},x)) \nonumber\\
\ &=&\frac{\gamma (x)}{2}|p(s,R_s^{t,r_0},x)-\overline{Z}%
^{p,t}(s)|^2-p(s,R_s^{t,r_0},x)\theta (s,R_s^{t,r_0}) \nonumber\\
\ &=&\frac{\gamma (x)}{2}\left| p(s,R_s^{t,r_0},x)-\overline{Z}^{p,t}(s)-\frac{%
\theta (s,R_s^{t,r_0})}{\gamma (x)}\right| ^2 \nonumber \\
&&-\overline{Z}^{p,t}(s)\theta (s,R_s^{t,r_0})-\frac{|\theta
(s,R_s^{t,r_0})|^2}{2\gamma (x)}.
\end{eqnarray}
By using the similar ideas as in Hu et al \cite{HIM2005} (see also
\cite{AID2010}) we deduce that
\begin{eqnarray}\label{5}
\overline{\pi }(s,R_s^{t,r_0},x)=\Psi_\Gamma \left( \overline{Z}^{\overline{p%
},t}(s)+\frac{\theta (s,R_s^{t,r_0})}{\gamma (x)}\right) \beta
^{-1}(s,R_s^{t,r_0}),
\end{eqnarray}
where
\begin{eqnarray*}
Y^{\overline{p},t}(r) &=&-x+\int_r^T\frac{\gamma (x)}2\text{\rm
dist}^2\left(
\overline{Z}^{\overline{p},t}(s)+\frac{\theta (s,R_s^{t,r_0})}{\gamma (x)}%
,\Gamma _s(\omega )\right) ds \\
&&\ -\int_r^T\left[ \overline{Z}^{\overline{p},t}(s)\theta (s,R_s^{t,r_0})+%
\frac{|\theta (s,R_s^{t,r_0})|^2}{2\gamma (x)}\right] ds-\int_r^TZ^{%
\overline{p},t}(s)dW(s),
\end{eqnarray*}
with $r\in [t,T].$ By substituting (\ref{5}) into (\ref{4}), we
further have for any $p\in \Gamma ,$
\[
\overline{g}(s,x,\overline{Z}^{\overline{p},t}(s),\overline{p}%
(s,R_s^{t,r_0},x))\leq \overline{g}(s,x,\overline{Z}%
^{p,t}(s),p(s,R_s^{t,r_0},x)),
\]
which means
\[
\overline{g}(s,x,\overline{Z}^{\overline{p},t}(s),\overline{p}%
(s,R_s^{t,r_0},x))=\mathop{\rm essinf}_p\overline{g}(s,x,\overline{Z}%
^{p,t}(s),p(s,R_s^{t,r_0},x)),
\]
is a necessary condition for $\overline{p}$ being optimal.
\end{example}
\begin{remark}Note that from (\ref{5}) we can tell the influence of
initial wealth on optimal strategy. For example, suppose $\alpha $
and $\beta $ are independent of $R_s^{t,r_0},$ $\alpha $ is
positive, we are aim to find an optimal strategy with no-shorting.
Hence (\ref{5}) becomes
\[
\overline{\pi }(s,x)=\Pi _{\Bbb{R}^{+}}\left( \frac{\theta (s)}{\gamma (x)}%
\right) \beta ^{-1}(s),\text{ }s\in [t,T],\text{ }x\in
\Bbb{R}\text{.}
\]
Since $\gamma $ is decreasing in $x,$ hence $\overline{\pi }(s,x)$
is increasing in $x,$ which means that the more initial wealth, the
more investment for the investor.
\end{remark}
\begin{remark}
Here we can not treat the general case of $g$ being quadratic growth
in $z$. One important reason for that lies in the limitation of
existed comparison theorem for quadratic BSDEs (see for example
Briand and Hu \cite{BH2008}) which is even not applicable on
deriving optimal strategy in Example 3.1. For example, some bounded
conditions of diffusion term in forward equation are indeed required
when studying quadratic Markovian BSDEs, see \cite{BH2008}.
\end{remark}
From Theorem 3.1 and Example 3.1 we can see the importance and
necessity of optimization problem for
\begin{eqnarray}\label{6}
\mathop{\rm essinf}_p\overline{g}(s,x,z,p(s,R_s^{t,r_0},x)),\text{ }s\in [t,T],\text{ }%
x,z\in \Bbb{R.}
\end{eqnarray}
Next we will study problem in (\ref{6}) from some new perspectives.
If $g$ is convex, we have
\begin{eqnarray*}
&&\ \mathop{\rm essinf}_p\left[ g(s,x,z-p(s,R_s^{t,r_0},x))+\theta
(s,R_s^{t,r_0})(z-p(s,R_s^{t,r_0},x))\right] \\
\ &=&-\mathop{\rm esssup}_p\left[ -\theta
(s,R_s^{t,r_0})(z-p(s,R_s^{t,r_0},x))-g(s,x,z-p(s,R_s^{t,r_0},x))\right],
\end{eqnarray*}
therefore
\begin{eqnarray}\label{7}
\mathop{\rm
essinf}_p\overline{g}(s,x,z,p(s,R_s^{t,r_0},x))=-G(s,x,\theta
(s,R_s^{t,r_0}))-\theta (s,R_s^{t,r_0})z,
\end{eqnarray}
where
\begin{eqnarray}\label{7.1}
G(s,x,\mu )=\sup_{r\in \text{Dom}(g)}\left( -\mu \cdot r-g(s,x,r)\right) ,%
\text{ }\mu \in \Bbb{R},\text{ }
\end{eqnarray}
with Dom$(g(s,x,\cdot ))=\{r\in \Bbb{R}:g(s,x,r)<+\infty \}$ being
the Legendre-Fenchel transform of $g(s,x,\cdot )$. We call
$G(s,x,\mu )$ the polar function. Once we get an optimal point of
(\ref{7.1}) $\overline{r}(s,R_s^{t,r_0},x)$, then
\begin{eqnarray}\label{8}
\overline{p}(s,R_s^{t,r_0},x)=z-\overline{r}(s,R_s^{t,r_0},x),\quad
s\in[t,T],\ x\in\mathbb{R},
\end{eqnarray}
is the optimal solution of problem (\ref{6}).
\begin{example}
\rm
 VaR (or AVaR) is currently one of most widely used
financial risk measures. It is shown that the solution $Y(\cdot )$
of BSDE with simple generator
\[
g(s,x,z)=\left\{
\begin{array}{c}
|z|, \\
-\frac{(\alpha -1)}\alpha |z|
\end{array}
\right.
\begin{array}{c}
\alpha <\frac 12, \\
\frac 12\leq \alpha \leq 1,
\end{array}
\]
is the limit of discrete AVaR under suitable conditions, see Section
7 in \cite{S2010}. If we consider problem (\ref{7.1}), we then have
\[
G(s,x,\mu )=\sup_{r\in \Bbb{R}}\left( -\mu \cdot r-|r|I_{\alpha <\frac 12}+%
\frac{(\alpha -1)}\alpha |r|I_{\alpha \in [\frac 12,1]}\right) ,\text{}%
\mu \in \Bbb{R}.
\]
After some calculations,
\[
G(s,x,\mu )=\left\{
\begin{array}{c}
0 \\
+\infty
\end{array}
\right.
\begin{array}{c}
\mu \in [-1,0], \\
\text{\rm others},
\end{array}
\]
and $\overline{r}=0$ is the optimal point of problem (\ref{7.1}).
\end{example}
\begin{remark}If $g(s,x,z)$ is Lipschitz in $z$, mathematically $\theta
(s;R_s^{t,r_0})$ should be less than the Lipschitz constant so as to ensure $%
G$ is finite, see p.35 in \cite{EPQ1997}. Moreover, such property
can have some more deeper and interesting financial interpretation
in some special cases, see Section 4 next.
\end{remark}
Before studying (\ref{6}), for any $%
(s,x)\in [0,T]\times \Bbb{R}$, we need the notion of
sub-differential of a convex function $f$ defined on $\Bbb{R}$.
Given $a,$ it is a set and we denote it by $\partial f(s,x,a)$;
where
\begin{eqnarray*}
\partial f(s,x,a) &=&\left\{\theta ;f(s,x,b)-f(s,x,a)\geq -\theta (b-a),\forall
b\in \text{Dom(}f(s,x,\cdot ))\right\},
\end{eqnarray*}
and the elements are called sub-gradient of $f$ at $a.$ By Section E
in \cite{HL2001} we can derive the following result.
\begin{theorem}
If $g$ is a continuous and convex function in $r$, and $G(s,x,\mu )$
is its
polar function, then $\widehat{r}\in \partial G(s,x,\mu )$ if and only if $%
\widehat{r}$ is optimal for the following problem, that is,
\[
-G(s,x,\mu )=\inf_r(g(s,x,r)+\mu \cdot r)=g(s,x,\widehat{r})+\mu
\cdot \widehat{r}.
\]
\end{theorem}
Given $\theta (s,R_s^{t,r_0}),$ if $\partial G(s,x,\theta
(s,R_s^{t,r_0}))$ is non-empty, then the optimal solution
exists. According to Section E in \cite{HL2001}, if $\mu $ is in the interior of Dom$%
(G(s,x,\cdot )),$ then $\partial G(s,x,\mu )$ is non-empty. In
particular, if $G$ is finite with $\mu \in \Bbb{R}$, then $\partial
G(s,x,\mu )$ is nonempty for any $\mu .$ Note that in Example 3.2,
if $\mu \in (-1,0),$ we can find at least one point $0\in
\partial G(s,x,\mu )$. Next we will give another example
when $G$ is finite.
\begin{example}
\rm
 Given $s$ and $x$, $g(s,x,r)=r^2$ with $r\in\mathbb{R}$; then
\[
G(s,x,\mu )=\sup_{r\in \Bbb{R}}(-\mu r-r^2),\text{ with }\mu \in
\Bbb{R}.
\]
After some calculations, $G(s,x,\mu )=\frac{\mu ^2}4$ and $\overline{r}%
=-\frac \mu 2$ is the optimal point. In this case, $G(s,x,\mu )$ is
differential, hence $\partial G(s,x,\mu )$ is nonempty for any $\mu \in \Bbb{%
R}$.
\end{example}
Next we will make use of (\ref{7}) above to derive the explicit form
of risk indifferent price, derivative hedge and marginal risk price.
Before it, we need a furthermore assumption.

 {\bf(H1)}
$F$ is bounded and differentiable with bounded derivative, $%
\frac{\partial \alpha (s,r)}{\partial r}$ and $\frac{\partial \beta (s,r)}{%
\partial r}$ are bounded.

Given $t\in[0,T]$, the risk indifference price $q$,
$\widehat{p}(\cdot ,R^{t,r_0}(\cdot ),x)=\widehat{\pi }(\cdot
,x)\beta (\cdot ,R^{t,r_0}(\cdot ))$, $\widehat{\pi }$ and
$\overline{\pi } $ being defined respectively in Section 2, we
derive that $X_{x-q}^{\widehat{\pi }}(T)$ is terminal wealth with
initial wealth being $x-q$, therefore, by (\ref{2.1}) and
(\ref{2.2}),
\begin{eqnarray*}
&&\ \ \ \ \ \rho _t(X_{x-q}^{\widehat{\pi }}(T)+F(R_T^{t,r_0}))=
\rho _t(X_{x-q}^{\widehat{p}}(T)+F(R_T^{t,r_0}))\\
\  &=&-X_{x-q}^{\widehat{p}}(T)-F(R_T^{t,r_0})+\int_t^Tg(s,x,Z^{\widehat{p%
},t}(s))ds-\int_t^TZ^{\widehat{p},t}(s)dW(s).
\end{eqnarray*}
Using similar ideas as above, from (\ref{7}) we have
\begin{eqnarray*}
&&\ \ \ \ \ \ \rho _t(X_{x-q}^{\widehat{\pi }}(T)+F(R_T^{t,r_0})) \\
\  &=&-x+q-F(R_T^{t,r_0})-\int_t^TG(s,x,\theta (s,R_s^{t,r_0}))ds \\
&&\ \ \ \ \ \ -\int_t^T\overline{Z}^{\widehat{p},t}(s)\theta
(s,R_s^{t,r_0})ds-\int_t^T\overline{Z}^{\widehat{p},t}(s)dW(s).
\end{eqnarray*}
Similarly
\begin{eqnarray*}
\rho _t(X_x^{\overline{\pi }}(T)) &=&-x-\int_t^TG(s,x,\theta
(s,R_s^{t,r_0}))ds \\
&&\ \ \ \ \ \ -\int_t^T\overline{Z}^{\overline{p},t}(s)\theta
(s,R_s^{t,r_0})ds-\int_t^T\overline{Z}^{\overline{p},t}(s)dW(s).
\end{eqnarray*}
As a result, the risk indifferent price $q\doteq q(t,x)$ can be
expressed as
\begin{eqnarray}\label{9}
q_t=\mathbb{E}_{\mathbb{Q}}^{\mathcal{F}_t}F(R_T^{t,r_0})=\frac{\mathbb{E}^{\mathcal{F}%
_t}A(T)F(R_T^{t,r_0})}{A(t)},
\end{eqnarray}
where $A(\cdot )$ is defined in $[t,T],$
\begin{eqnarray}\label{10}
\left.\frac{d\mathbb{Q}}{d\mathbb{P}}\right|_{\mathcal{F}_t}=A(s)=e^{-\int_t^s\theta
(v,R_v^{t,r_0})dW(v)-\frac 12\int_t^s\theta
^2(v,R_v^{t,r_0})dv},\text{ }s\in [t,T].
\end{eqnarray}
Note that the indifference price $q$ is linear in $F(R^{t,r_0}_T)$
and independent of $x,$ therefore the marginal risk price of
derivative is still $q_t$ itself, i.e., $p(t,x)=q(t,x)$. As to the
derivative hedge, it follows from (\ref{8}) that
\[
\Delta (s,R_s^{t,r})=\frac{[\widehat{p}(s,R_s^{t,r_0},x)-\overline{p}%
(s,R_s^{t,r_0},x)]}{\beta (s,R_s^{t,r_0})}=\frac{[\overline{Z}^{\widehat{p}%
,t}(s)-\overline{Z}^{\overline{p},t}(s)]}{\beta (s,R_s^{t,r_0})},
\]
with $s\in [t,T].$ If we denote
\[
\Delta \overline{Y}^t(r)=\overline{Y}^{\widehat{p},t}(r)-\overline{Y}^{%
\overline{p},t}(r),\text{ }\Delta \overline{Z}^t(r)=\overline{Z}^{\widehat{p}%
,t}(r)-\overline{Z}^{\overline{p},t}(r),\text{ }r\in [t,T],
\]
where for example
\begin{eqnarray*}
\overline{Y}^{\widehat{p},t}(r)
&=&-x-F(R_T^{t,r_0})-\int_t^TG(s,x,\theta
(s,R_s^{t,r_0}))ds \\
&&\ -\int_t^T\overline{Z}^{\widehat{p},t}(s)\theta (s,R_s^{t,r_0})ds-\int_t^T%
\overline{Z}^{\widehat{p},t}(s)dW(s),
\end{eqnarray*}
then
\begin{eqnarray*}
\Delta \overline{Y}^t(r) &=&-F(R_T^{t,r_0})-\int_r^T\Delta \overline{Z}%
^t(s)\theta (s,R_s^{t,r_0})ds-\int_r^T\Delta \overline{Z}^t(s)dW(s) \\
\  &=&-F(R_T^{t,r_0})-\int_r^T\Delta
\overline{Z}^t(s)d\widehat{W}(s),
\end{eqnarray*}
where
\[
\widehat{W}(s)=\int_t^s\theta (v,R_v^{t,r_0})dv+W(s),\text{ }s\in
[t,T].
\]
By using {\bf (H1)}, we can verify the results of (see the appendix
for detailed proof),
\begin{eqnarray}\label{10.1}
&&\mathbb{E}_{\Bbb{Q}}|F(R_T^{t,r_0})| <\infty ,\text{
}\mathbb{E}_{\Bbb{Q}}\left[
\int_t^T|D_sF(R_T^{t,r_0})|^2ds\right] ^{\frac 12}ds<\infty , \nonumber \\
&&\mathbb{E}_{\Bbb{Q}}\left[ |F(R_T^{t,r_0})|\cdot \left(
\int_t^Th(r)dr\right) ^{\frac 12}\right] <\infty ,
\end{eqnarray}
with $D_sM$ being the Malliavin derivative of random variable $M$,
and
\[
h(r)=\left( \int_t^TD_r\theta (s,R_s^{t,r_0})dW(s)+\int_t^TD_r\theta
(s,R_s^{t,r_0})\cdot \theta (s,R_s^{t,r_0})ds\right) ^2,
\]
therefore we have
\begin{eqnarray}\label{11}
\Delta Z^t(r)=-\mathbb{E}_{\mathbb{Q}}^{\mathcal{F}_r}\left[
D_rF(R_T^{t,r_0})-F(R_T^{t,r_0})\int_r^TD_r\theta (s,R_s^{t,r_0})d\widehat{W}%
(s)\right] ,
\end{eqnarray}
therefore the derivative hedge $\Delta (s,R_s^{t,r})$ with $s\in
[t,T]$ can be given by
\begin{eqnarray}\label{11.1}
\Delta (s,R_s^{t,r_0}) &=&-\frac{\mathbb{E}^{\mathcal{F}_s}\left[
A(T)F_r(R_T^{t,r_0})\cdot D_sR_T^{t,r_0}\right] }{\beta
(s,R_s^{t,r_0})A(s)}
\\
&&\ +\frac{\mathbb{E}^{\mathcal{F}_s}\left[
A(T)F(R_T^{t,r_0})\int_s^T\theta
_r(u,R_u^{t,r_0})D_sR_u^{t,r_0}d\widehat{W}(u)\right] }{\beta
(s,R_s^{t,r_0})A(s)}.\nonumber
\end{eqnarray}
To sum up, we have
\begin{theorem}
Suppose {\bf(H1)} hold, $g$ is Lipschitz in $z$, then the
indifference price $q(t,x)$ and the derivative hedge
$\Delta(s,R^{t,r_0}_s)$ can be expressed in (\ref{9}) and
(\ref{11.1}) respectively. In addition, $p(t,x)=u(t,x)$.
\end{theorem}
\begin{remark}
Mathematically the expression of $p,$ $q$ and $\Delta$ just depends
on some parameters of financial market. Some other subjective
parameter represented by $g$, such as risk aversion parameter, will
not affect them. Since all the variable are supposed to be one
dimensional, $\beta$ is invertible, and admissible strategy is
non-constrained, one important reason for this phenomenon, we
believe, is the completeness of financial market. Otherwise, for
example, if the admissible strategy is restricted to take value in
subset of $\mathbb{R}$, then $q(t,x)$ will depend on $g$ and some
delicate analysis will be needed to get the explicit expression, see
\cite{AID2010}.
\end{remark}
\begin{remark}
As we know, in the real world there are lots of deals that do not go
through the exchange trading and people called them
over-the-counter(OTC for short). Hence how to find a fair and
suitable price for the agreements between the sellers and buyers is
faced by certain investment institutions who are offering the OTC
deals. Of course, from the view of minimizing risk for financial
position, risk indifference price seems to be a reasonable choice.
Generally speaking, this price should be nonlinear on the amount of
derivatives (see \cite{AID2010}) which prompt us to introduce
marginal risk price to characterize the price for each one
derivative. However, here we are lucky to get the linear form, which
is to say, the price of $k\times F(R^{t,r_0}_T)$ is just $k$ times
the price of $F(R^{t,r_0}_T)$. This seems a little too ideal to have
"discount" in this procedure. We believe that the requirement of
complete financial market is also one of main reasons for that, and
we hope to show more results of incomplete financial market case in
future.
\end{remark}
\section{Some special risk minimization problems with explicit solutions}
In this section, we will give several special cases to illustrate
the well application of Theorem 3.1 above, where there are two basic
financial notions, market price of risk (see for example
\cite{KS1998}) and risk aversion parameter involved. The former one
is objective and determined by the financial market while the later
one is subjective and mainly depends on the investor. We will
consider their relations with the optimal strategy of our risk
minimization problem above.
\subsection{Case 1}
In this subsection, we consider one risk minimization problem by
supposing
\[
g(s;z,x)=k(s;R_s^{t,r_0};x)\left( \sqrt{1+|z|^2}-1\right) ,\text{
}s\in [t,T],\text{ }x,z\in \Bbb{R},
\]
where $k(s;r;x)$: $[0;T]\times \Bbb{R}\times \Bbb{R}\rightarrow
\Bbb{R}$ is positive, decreasing in $x$; and $k(s;r;x)>|\theta
(s;r)|$ with $s\in [0;T]$ and $r\in \Bbb{R};$ $x\in \Bbb{R}$.
Suppose that

We claim that $k$ above can be regarded as a subjective parameter
representing investors' attitude towards risk, that is, the larger
$k$ becomes; the more risk averse the investor is. Actually, suppose
that two investors $A$ and $B$ are facing one contingent claim
$\xi$, and they will measure its risk in two different way, such as
two BSDEs with generator $g_1$ and $g_2$ respectively,
\[
g_i(s;z,x)=k_i(s;R_s^{t,r_0};x)\left( \sqrt{1+|z|^2}-1\right)
,\text{ }s\in [t,T],\text{ }x,z\in \Bbb{R},\text{ }i=1,2.
\]
In addition, $k_1\leq k_2.$ It follows from the comparison theorem
for BSDEs that $\rho _t^A(\xi )\leq \rho _t^B(\xi )$ which means
that the risk considered by investor $A$ is less than that by $B$.
In other words, investor $A$ is less sensitive to risk than $B$, or
is more risk tolerance (less risk averse) than $B$ when facing the
same uncertainty. This can be identified as one important role of
$k$. Next we will derive the explicit expression of optimal strategy
according to Theorem 3.1. First we arrive at,
\begin{eqnarray*}
&&\overline{g}(s,x,\overline{Z}^{p,t}(s),p(s,R_s^{t,r_0},x)) \\
&=&k(s,R_s^{t,r_0},x)\left( \sqrt{1+|\overline{Z}%
^{p,t}(s)-p(s,R_s^{t,r_0},x)|^2}-1\right) -p(s,R_s^{t,r_0},x)\theta
(s,R_s^{t,r_0}),
\end{eqnarray*}
and
\begin{eqnarray*}
&&\frac{\partial \overline{g}(s,x,\overline{Z}^{p,t}(s),p(s,R_s^{t,r_0}))}{%
\partial p} \\
&=&k(s,R_s^{t,r_0},x)\frac{p(s,R_s^{t,r_0},x)-\overline{Z}^{p,t}(s)}{\sqrt{%
1+|p(s,R_s^{t,r_0},x)-\overline{Z}^{p,t}(s)|^2}}-\theta
(s,R_s^{t,r_0}).
\end{eqnarray*}
Therefore, if $-1<\frac{\theta
(s,R_s^{t,r_0})}{k(s,R_s^{t,r_0},x)}<0,$ by Theorem 3.1, there
exists a unique optimal strategy $\overline{p}(s,R_s^{t,r_0},x)$,
together with $\overline{Z}^{\overline{p},t}(s)$ such that
$\frac{\partial
\overline{g}(s,x,\overline{Z}^{\overline{p},t}(s),\overline{p}%
(s,R_s^{t,r_0},x))}{\partial p}=0,$ where
\begin{eqnarray}\label{12}
\overline{p}(s,R_s^{t,r_0},x)=\frac{\theta (s,R_s^{t,r_0})}{\sqrt{%
k^2(s,R_s^{t,r_0},x)-\theta ^2(s,R_s^{t,r_0})}}+\overline{Z}^{\overline{p}%
,t}(s),
\end{eqnarray}
and
\begin{eqnarray}\label{13}
\overline{Y}^{\overline{p},t}(r) &=&-x+\int_r^T\left[ \sqrt{%
k^2(s,R_s^{t,r_0},x)-\theta
^2(s,R_s^{t,r_0})}-k(s,R_s^{t,r_0},x)\right] ds \nonumber
\\
&&\ -\int_r^T\overline{Z}^{\overline{p},t}(s)\theta
(s,R_s^{t,r_0})ds-\int_r^T\overline{Z}^{\overline{p},t}(s)dW(s),
\end{eqnarray}
is the so-called minimal risk value at time $r\geq t.$ By using
Girsanov theorem, we can rewrite (\ref{13}) into
\[
\overline{Y}^{\overline{p},t}(r)=-x+\mathbb{E}_{\Bbb{Q}}^{\mathcal{F}%
_t}\int_r^T\left[ \sqrt{k^2(s,R_s^{t,r_0},x)-\theta ^2(s,R_s^{t,r_0})}%
-k(s,R_s^{t,r_0},x)\right] ds,
\]
where $A(\cdot)$ is defined in (\ref{10}) and
\[
d\Bbb{Q}=A(T)d\Bbb{P},\text{ }s\in [t,T].
\]
If $0<\frac{\theta (s,R_s^{t,r_0})}{k(s,R_s^{t,r_0},x)}<1,$ we can
also prove that the above $\overline{p}$ is optimal. Particularly,
when $\theta $ and $k$ are deterministic, i.e., they are independent
of $R_s^{t,r_0},$ then
\[
\overline{Z}^{\overline{p},t}(s)=0,\text{ }\overline{p}(s,x)=\frac{\theta (s)%
}{\sqrt{k^2(s,x)-\theta ^2(s)}}\text{ with }s\in [t,T],
\]
hence the optimal strategy
\begin{eqnarray} \label{13.1}
\overline{\pi }(s,x)=\frac{\theta (s)}{\sqrt{k^2(s,x)-\theta
^2(s)}}\beta ^{-1}(s),
\end{eqnarray}
 and minimal risk value at time
$t$ is described by
\[
\overline{Y}^{\overline{p}}(t)=-x+\int_t^T[\sqrt{k^2(s,x)-\theta ^2(s)}%
-k(s,x)]ds.
\]
If $\beta(\cdot)$ is positive, then from (\ref{13.1}) the optimal
strategy $\overline{\pi}$ is increasing in market price of risk.
Similar point also appeared by Pirvu and Zhang in \cite{PZ2012}
where optimal investment, consumption and life insurance acquisition
for wage earners with constant relative risk aversion (CRRA)
preference is carried out. On the other hand, (\ref{13.1}) also
indicates that $\overline{\pi}$ is decreasing in $k$. By combining
this to the role of $k$ claimed above, it is easy to say, the larger
$k$ becomes, the more risk averse the investor is, and the less
investment in the risky asset, which is consistent with results in
\cite{X2011}.
\begin{remark}
Let us give two points on parameter $k$ here. Inspired by arguments
in \cite{BMZ2012} and \cite{HJZ2012}, we suppose that risk aversion
parameter $k$ is dependent on $x$. Actually, if $k$ is independent
of $x$, from (\ref{12}) and (\ref{13}) $p$ is also independent of
$x.$ This means that optimal strategy is independent of the initial
wealth, which seems unreasonable according to the arguments in
Section 3 of \cite{BMZ2012}. On the other hand, $k$ is also supposed
to be decreased in $x$ in order to make the results more meaningful.
For example, by (\ref{13.1}) $\overline{\pi }(s,x)$ is decreasing in $%
k$. Since $\alpha$ represents the appreciation rates for the risky
asset, it is
suitable to suppose that $\alpha >0$. Hence $%
\overline{\pi }(s;x)$ is increasing in $x$, which fits the common
economic knowledge. After all, one usually can not invest less money
in the risky when one's initial wealth is 100,000,000 dollars than
one would if the initial wealth is 100 dollars. Such a relation
between optimal strategy and initial wealth also hold in utility
maximization problem with utility function $U$ exhibiting decreasing
absolute risk aversion (DARA for short), see for example
\cite{B2007}.
\end{remark}
\begin{remark}\label{re:4.3}
As we can see from (\ref{13.1}), $\overline{\pi}$ is increasing
(decreasing) in $\theta$ when $\beta$ is positive (negative).
However, in order to make the risk be minimized, there should be a
threshold for $\theta$, which for example means people can not keep
on investing more all the time when $\theta$ increases and $\beta$
is positive. This indicates that for a given investor with
subjective parameter $k$, his/her investment strategy $\theta$ is
controlled by $k$. From the view of mathematics, the case of
$\left|\frac{\theta (s,R_s^{t,r_0})}{k(s,R_s^{t,r_0},x)}\right|> 1$
will cause the illposedness of optimal strategy. On the other hand,
given two investors $A$ (with $k_1$) and $B$ (with $k_2$) and
$k_1\leq k_2$, $\left|\theta(s,R_s^{t,r_0})\right|\leq k_1$ implies
that $\left|\theta(s,R_s^{t,r_0})\right|\leq k_2$, which means
investor $B$ has larger boundary to keep market price of risk. It
also indicates that people who are more risk aversion have larger
probability to find the optimal strategy to minimize the risk. Such
relations among subjective parameter (like $k$), objective parameter
(like $\theta$), and wellposedness of risk minimization problem are
new to our best knowledge.
\end{remark}
\begin{remark}
Here the parameter $k$ is assumed to keep unchanged as the time goes
by. However, in real world, people's attitude to risk and
uncertainty is always different at different time period. For
example, as is shown in \cite{PZ2012}, when the wage earner becomes
older he/she has a higher demand for life insurance/pension
annuities due to the higher hazard rate. That is to say, a more
general but applicable case for $k$ should be $k(t;s;r;x)$. In this
case, the risk is described by
\begin{eqnarray*}
&&\overline{Y}^{p,t}(t,r) \\
&=&-x+\int_r^Tk(t,s,R_s^{t,r_0},x)\left( \sqrt{1+|\overline{Z}%
^{p,t}(t,s)-p(s,R_s^{t,r_0},x)|^2}-1\right) ds \\
&&-\int_r^T[p(s,R_s^{t,r_0},x)\theta (s,R_s^{t,r_0})-\overline{Z}%
^{p,t}(t,s)\theta
(s,R_s^{t,r_0})]ds-\int_r^T\overline{Z}^{p,t}(t,s)dW(s),
\end{eqnarray*}
with $r\in [t,T],$ and $\overline{p},$ depending on $t,$ should be
\begin{eqnarray}\label{13.2}
\overline{p}(t,s,R_s^{t,r_0},x)=\frac{\theta (s,R_s^{t,r_0})}{\sqrt{%
k^2(t,s,R_s^{t,r_0},x)-\theta ^2(s,R_s^{t,r_0})}}+\overline{Z}^{\overline{p}%
,t}(t,s),
\end{eqnarray}
with $s\in [t,T],$ where $\overline{Z}^{\overline{p},t}(t,s)$
satisfies
\begin{eqnarray*}
\overline{Y}^{p,t}(t,r) &=&-x+\int_r^T\left( \sqrt{k^2(t,s,R_s^{t,r_0},x)-%
\theta ^2(s,R_s^{t,r_0})}-k(t,s,R_s^{t,r_0},x)\right) ds \\
&&\ \ -\int_r^T\overline{Z}^{p,t}(t,s)\theta (s,R_s^{t,r_0})ds-\int_r^T%
\overline{Z}^{p,t}(t,s)dW(s).
\end{eqnarray*}
If $r=t$, $(\varphi ^{\overline{p},t}(\cdot ),\overline{Z}^{\overline{p}%
,t}(\cdot ,\cdot ))$ satisfies a linear backward stochastic Volterra
integral equation (BSVIE for short) studied for example in \cite{Y2008}, with $\varphi ^{\overline{p},t}(t)=%
\overline{Y}^{\overline{p},t}(t,t).$ From (\ref{13.2}) $\pi$ is not
time consistent strategy due to dependence on time $t.$ At the
moment, it is not clear to us how one can study the time consistent
solution in such case.
\end{remark}

\subsection{Case 2}

In this subsection, we consider one special case of {\bf
Problem(RM)} by supposing
\[
g(s,x,z)=l(s;R_s^{t,r_0},x)\cdot \text{ln}\frac 12(1+e^{-z}),\text{
}s\in [t,T],\text{ }x,z\in \Bbb{R},
\]
where $l(s;r;x):[0;T]$ $\times \Bbb{R}$ $\times $ $\Bbb{R}$
$\rightarrow \Bbb{R}$ is bounded and positive. By the similar
analysis as above, $l$ is also a subjective parameter for investors,
that is, the larger $l$ becomes, the more risk averse the investor
is. In order to derive explicit strategy in such setting, firstly we
have
\begin{eqnarray*}
&&\overline{g}(s,x,\overline{Z}^{p,t}(s),p(s,R_s^{t,r_0},x)) \\
&=&l(s,R_s^{t,r_0},x)\ln \frac{1+\exp \left\{p(s,R_s^{t,r_0})-\overline{Z}%
^{p,t}(s)\right\}}2-p(s,R_s^{t,r_0},x)\theta (s,R_s^{t,r_0}),
\end{eqnarray*}
and
\begin{eqnarray*}
&&\frac{\partial \overline{g}(s,x,\overline{Z}^{p,t}(s),p(s,R_s^{t,r_0},x))}{%
\partial p} \\
&=&l(s,R_s^{t,r_0},x)\frac{\exp \left\{p(s,R_s^{t,r_0},x)-\overline{Z}^{p,t}(s)\right\}%
}{1+\exp
\left\{p(s,R_s^{t,r_0},x)-\overline{Z}^{p,t}(s)\right\}}-\theta
(s,R_s^{t,r_0}).
\end{eqnarray*}
If $\frac{\theta (s,R_s^{t,r_0})}{l(s,R_s^{t,r_0},x)}<0,$ i.e.,
$\theta
(s,R_s^{t,r_0})$ is negative, then $\frac{\partial \overline{g}(s,x,%
\overline{Z}^{p,t}(s),p(s,R_s^{t,r_0},x))}{\partial p}>0,$ therefore
the
optimal strategy does not exist. If $\frac{\theta (s,R_s^{t,r_0})}{%
l(s,R_s^{t,r_0},x)}\geq 1,$ then $\frac{\partial \overline{g}(s,x,\overline{Z%
}^{p,t}(s),p(s,R_s^{t,r_0},x))}{\partial p}<0,$ hence the optimal
strategy also does not exist.

As to the case of $0\leq \theta (s,R_s^{t,r_0})\leq
l(s,R_s^{t,r_0},x),$ by Theorem 3.1 there exists only one
$\overline{p}(s,R_s^{t,r_0},x)$
such that $\frac{\partial \overline{g}(s,x,\overline{Z}^{\overline{p},t}(s),%
\overline{p}(s,R_s^{t,r_0},x))}{\partial p}=0,$ where
\[
\overline{p}(s,R_s^{t,r_0},x)=\ln \left[ \frac{\theta (s,R_s^{t,r_0})}{%
l(s,R_s^{t,r_0},x)-\theta (s,R_s^{t,r_0})}\right] +\overline{Z}^{\overline{p}%
,t}(s),
\]
$\ $and $\overline{Z}^{\overline{p},t}(s)$ satisfies
\begin{eqnarray*}
\overline{Y}^{\overline{p},t}(r) &=&-x+\int_r^Tl(s,R_s^{t,r_0},x)\ln \frac{%
l(s,R_s^{t,r_0},x)}{2[l(s,R_s^{t,r_0},x)-\theta (s,R_s^{t,r_0})]}ds \\
&&-\int_r^T\theta (s,R_s^{t,r_0})\ln \frac{\theta (s,R_s^{t,r_0})}{%
l(s,R_s^{t,r_0},x)+\theta (s,R_s^{t,r_0})}ds \\
&&-\int_r^T\overline{Z}^{\overline{p},t}(s)\theta (s,R_s^{t,r_0})ds-\int_r^T%
\overline{Z}^{\overline{p},t}(s)dW(s).
\end{eqnarray*}
Note that when $\theta $ and $l$ are deterministic,
\[
\overline{Z}^{\overline{p},t}(s)=0,\text{ }\overline{p}(s,x)=\ln
\left[ \frac{\theta (s)}{l(s,x)-\theta (s)}\right] \text{ with }s\in
[t,T],
\]
hence the optimal strategy
\begin{eqnarray}\label{14}
\overline{\pi }(s)=\ln \left[ \frac{\theta (s)}{l(s,x)-\theta
(s)}\right] \beta ^{-1}(s),
\end{eqnarray}
and minimal risk value at time $t$ is described by
\[
\overline{Y}^{\overline{p}}(t)=-x+\int_t^T\left( l(s,x)\ln \frac{l(s,x)}{%
2[l(s,x)-\theta (s)]}-\theta (s)\ln \frac{\theta (s)}{l(s,x)-\theta (s)}%
\right) ds.
\]
Suppose $\alpha $ is non-negative, $\theta $ is positive, then
$\beta $ is also positive. Comparing with Case 1, (\ref{14}) also
indicates three points which are consistent with the one in
\cite{B2007}, \cite{PZ2012} and \cite{X2011}. Firstly the optimal
strategy $\overline{\pi}$ is increasing in market price of risk
$\theta$. Secondly, $\overline{\pi}$ is decreasing in $l$, i.e., the
more risk averse investor is, the less investment he/she will spend
in risky assets. Thirdly, since $l$ is decreasing in initial wealth
$x$, by combining the second point above, $\overline{\pi}$ is
increasing in $x$ which reflects the relations between optimal
strategy and initial wealth.
\begin{remark}
Speaking of the comparison of results between Case 1 and here, on
the one hand, similar ideas in the remarks there are also applicable
here. On the other hand, in order to ensure the existence of optimal
strategy, the investor with such risk preference has more
requirements (or restriction) of financial market. For example,
market price of risk here should be assumed to be non-negative,
otherwise the more investment in the risky asset, the more risk
he/she will take, which means there is no best choice for him/her.
Since it is reasonable to have $\alpha>0$, then positive
$\overline{\pi}$ implies positive volatility rate of risky asset
$\beta$ which, in the literature, is not an unacceptable condition
on financial market. Also negative $\beta$ in some sense reflects
certain negative (or unexpected) influence of random factors on the
price of risky asset, which reasonably implies the increasing risk
for investor.
\end{remark}
\subsection{Case 3}
In this case, we will consider one special case of {\bf Problem(RM)}
by supposing generator to be
\begin{eqnarray}\label{15.0}
 g(x,z)=\left\{
\begin{array}{c}
|z|-\frac 1{2\gamma (x)}, \\
\frac{\gamma (x)}2|z|^2,
\end{array}
\right.
\begin{array}{c}
|z|\geq \frac 1{\gamma (x)}, \\
|z|<\frac 1{\gamma (x)},
\end{array}
\end{eqnarray}
with $\gamma (x)$ being a decreasing positive function. Note that $%
h(x,z)=\gamma (x)g(x,z)$ is the Huber penalty function (see
\cite{H1964}) which plays an important role in robust statistics. As
proved in \cite{S2009} and \cite{S2010}, such BSDE is regarded as
the continuous time analogue of the discrete Gini principle of form
\[
V^\gamma (t_i,\xi )=\mathop{\rm essinf}_{\Bbb{Q}}\left(
\mathbb{E}_{\Bbb{Q}}^{\mathcal{F}_t}\xi
+\frac 1{2\gamma (x)}C_{t_i}(\left. \Bbb{Q}\right| \Bbb{P})\right) ,\text{ }%
t_i\in [0,T],
\]
with $C_{t_i}(\left. \Bbb{Q}\right| \Bbb{P})$ being the Gini index
(we refer the reader to see the meaning of notations there). From
(\ref{15.0}), $g$ is increasing in $\gamma,$ thus $\gamma$, like $k$
and $l$ before, can also represent investor's attitude to risk. On
other hand, according to Theorem 24 in \cite{MMR2006}, under certain
conditions imposed on $\xi$, Gini principle of $V^\gamma (0,\xi )$
agrees with mean-variance functional of the form
\begin{eqnarray}\label{15}
J(0,\xi )=\mathbb{E}\xi -\frac 1{2\gamma (x)}\text{Var(}\xi ),
\end{eqnarray}
where $\gamma$ in the mean-variance functional (\ref{15}) is a risk
aversion parameter, see for example \cite{BMZ2012}. Therefore,
roughly speaking, it follows from above two points that $\gamma$ in
(\ref{15.0}) is the risk aversion parameter in our setting.

Next we will derive the expression of optimal strategy by means of
Theorem 3.1. When $|\theta (s,R_s)|\leq 1$; we have
\begin{eqnarray*}
\overline{g}(x,z,p) &=&\left\{
\begin{array}{c}
z-p-p\theta -\frac 1{2\gamma (x)}, \\
p-z-p\theta -\frac 1{2\gamma (x)}, \\
\frac{\gamma (x)}2|p-z|^2-p\theta ,
\end{array}
\right.
\begin{array}{c}
z-p\geq \frac 1{\gamma (x)}, \\
p-z\geq \frac 1{\gamma (x)}, \\
-\frac 1{\gamma (x)}\leq p-z\leq \frac 1{\gamma (x)}
\end{array}
\\
&\geq &\left\{
\begin{array}{c}
-z\theta +\frac 1{2\gamma (x)}+\frac \theta {\gamma (x)}, \\
-z\theta +\frac 1{2\gamma (x)}-\frac \theta {\gamma (x)}, \\
-z\theta -\frac{\theta ^2}{2\gamma (x)},
\end{array}
\right.
\begin{array}{c}
z-p\geq \frac 1{\gamma (x)}, \\
p-z\geq \frac 1{\gamma (x)}, \\
-\frac 1{\gamma (x)}\leq p-z\leq \frac 1{\gamma (x)}.
\end{array}
\end{eqnarray*}
After some basic calculations, by using Theorem 3.1, we arrive at
\begin{eqnarray}\label{16}
\overline{p}(s,R_s^{t,r_0},x)=\overline{Z}^{\overline{p},t}(s)+\frac{\theta
(s,R_s^{t,r_0})}{\gamma (x)},
\end{eqnarray}
where
\[
\overline{Y}^{\overline{p}}(r)=-x-\int_r^T\left( \overline{Z}^{\overline{p}%
,t}(s)\theta (s,R_s^{t,r_0})+\frac{\theta ^2(s,R_s^{t,r_0})}{2\gamma (x)}%
\right) ds-\int_t^T\overline{Z}^{\overline{p},t}(s)dW(s),
\]
with $r\in [t,T].$ As to the case of $|\theta (s,R_s^{t,r_0})|>1,$
we have $\inf\limits_p\overline{g}(x,z,p)=-\infty,$ which means
optimal strategy does not exist.
\begin{remark}
Since $\gamma $ in (\ref{15.0}) can be seen as risk aversion
parameter, it is natural to be decreasing in initial wealth from a
obvious intuition. To study the relation between initial wealth and
optimal strategy, we look at a special case, i.e., $\theta$ is
independent of $R^{t,r_0}_s.$ By (\ref{16}), we have
\begin{eqnarray}\label{17}
\overline{\pi }(s;x)=\frac{\alpha (s)}{\beta ^2(s)\gamma (x)}, \quad
s\in[t,T],\ x\in\mathbb{R}.
\end{eqnarray}
By reasonably assuming $\alpha $ is non-negative, we deduce that $
\overline{\pi }(s;x)$ is increasing in initial wealth since $\gamma$
is decreasing in $x$. This is consistent with the results in Example
3.1, the above two cases and the utility maximization problem with
$U$ exhibiting decreasing absolute risk aversion in \cite{B2007}.
\end{remark}
\begin{remark}
By looking at (\ref{17}), and supposing $\alpha$ to be non-negative,
$\overline{\pi}$ is decreasing in risk aversion parameter $\gamma$,
which is a comparative result among investors consistent with the
one in Section 4 of \cite{X2011}.
\end{remark}
\begin{remark}
By assuming $\beta$ to be positive, from (\ref{17}), optimal
strategy $\overline{\pi}$ is also increasing in market price of risk
$\theta$ which is consistent with Case 1 and 2 above. The difference
between there and here is the boundary of $\theta$ in such setting
is a constant independent of certain subjective parameter.
\end{remark}

 Appendix

In this section, we will verify the conditions in (\ref{10.1}) in
order to use the generalized Clark representation formula in
\cite{OK1991}. For readers' convenience, we give several basic
notations required in the sequel. For given $t\in[0,T]$,
$\mathcal{S}[t,T]$ is the class of smooth random variable $F$ of
\[
F=f\left( \int_t^Th_1(s)dW(s),\cdots ,\int_t^Th_n(s)dW(s)\right) ,\text{ }%
f\in C_p^\infty (R^n),\text{ }h_i\in L^2[t,T],
\]
with $n\geq 1,$ $L^2[t,T]$ being the set of deterministic square
integral functions, $D_{r}H$ being the Malliavin derivative of $H$
at time $r$. We denote by $\Bbb{D}^{1,2}[t,T]$ the closure of the
class of smooth random variables $\mathcal{S}[t,T]$ with respect to
the norm
\[
||H||_{1,2}=\left[
\mathbb{E}|H|^2+\mathbb{E}\int_t^T|D_rH|^2dr\right] ^{\frac 12}.
\]
We denote by $\Bbb{L}^{1,2}[t,T]$ the class of square integral,
$\mathbb{F}$-adapted process $x$ such that $x(r)\in \Bbb{D}^{1,2}$
for almost $r\in [t,T],$ and
\[
\mathbb{E}\int_t^T\int_t^T|D_sx(t)|^2dtds<\infty ,
\]
with $D_{s}x$ being a measurable version. For given $r,s\in [t,T],$
since $\frac{\partial b(s,r)}{\partial r},$ $\frac{\partial \sigma (s,r)}{%
\partial r}$ exists and are continuous, by Theorem 2.2.1 in \cite{N2005}, we have $%
D_rR_s^{t,r_0}\in \Bbb{D}^{1,2}$,
\[
D_rR_s^{t,r_0}=\sigma
(r,R_r^{t,r_0})+\int_r^sb_r(u,R_u^{t,r_0})D_rR_u^{t,r_0}du+\int_r^s\sigma
_r(u,R_u^{t,r_0})D_rR_u^{t,r_0}dW(u),
\]
for $r\leq s,$ a.e., $D_rR_s^{t,r_0}=0$ for $r>s,$ and
\begin{eqnarray}\label{17}
\sup_{r\in [t,T]}\mathbb{E}\left( \sup_{s\in
[r,T]}|D_rR^{t,r_0}(s)|^k\right) <\infty ,\text{ with }k\geq 2.
\end{eqnarray}
For $r\leq s,$ $s\in[t,T],$ we can express $D_rR_s^{t,r_0}$ by
\[
D_rR_s^{t,r_0}=\sigma
(r,R_r^{t,r_0})e^{\int_r^sb_r(u,R_u^{t,r_0})du+\int_t^s\sigma
_r(u,R_u^{t,r_0})dW(u)-\frac 12\int_t^s\sigma
_r^2(u,R_u^{t,r_0})du}.
\]
Since $\frac{\partial \alpha (s,r)}{\partial r}$ and $%
\frac{\partial \beta (s,r)}{\partial r}$ are bounded, then
\[
\frac{\partial \theta (t,r)}{\partial r}=\frac 1{\beta
(t,r)}\frac{\partial \alpha (t,r)}{\partial r}-\frac{\alpha
(t,r)}{\beta ^2(t,r)}\frac{\partial \beta (t,r)}{\partial r}
\]
is bounded, thus by combining (\ref{17}), we have
\[
\mathbb{E}\int_t^T\int_t^T|D_r\theta (s,R_s^{t,r_0})|^2drds<\infty ,
\]
and we get that $\theta (s,\cdot )\in \Bbb{L}^{1,2}[t,T].$ Since $F$
has bounded derivative, we then get $F(R_T^{t,r_0})\in
\Bbb{D}^{1,2}[t,T].$ Next we will verify the condition in
(\ref{10.1}). Obviously the first inequality is easy to see. As to
the second one, we have
\begin{eqnarray*}
\mathbb{E}_Q\left[ \int_t^T|D_sF(R_T^{t,r_0})|^2ds\right] ^{\frac
12} &=&\mathbb{E}\left(
A(T)\left[ \int_t^T|D_sF(R_T^{t,r_0})|^2ds\right] ^{\frac 12}\right) \\
\ &\leq &C\left(\mathbb{E}A^2(T)\right) ^{\frac 12}\left(
\mathbb{E}\int_t^T|D_sF(R_T^{t,r_0})|^2ds\right) ^{\frac 12}.
\end{eqnarray*}
Since $\theta $ is bounded, thus $\mathbb{E}A^2(T)<\infty ,$ and
therefore
\[
\mathbb{E}\int_t^T|D_sF(R_T)|^2ds=\mathbb{E}\int_t^T|F_r(R_T)|^2|D_sR_T|^2ds<\infty
,
\]
due to the fact that $F$ has bounded derivative. As to the last one,
we have
\begin{eqnarray*}
&&\ \ \mathbb{E}\left( A(T)\left[ \int_t^T\left( \int_t^TD_u\theta
(s,R_s^{t,r_0})dW(s)\right) ^2du\right] ^{\frac 12}\right) \\
\ &\leq &C\left( \mathbb{E}A^2(T)\right) ^{\frac 12}\cdot \left[
\mathbb{E}\int_t^T\left(
\int_t^T|D_u\theta (s,R_s^{t,r_0})|^2ds\right) du\right] ^{\frac 12} \\
\ &\leq &C\left[ \mathbb{E}\int_t^T\left( \int_t^T\left|
\frac{\partial \theta (s,R_s^{t,r_0})}{\partial r}\right| ^2\left|
D_uR_s^{t,r_0}\right| ^2ds\right) du\right] ^{\frac 12}<\infty ,
\end{eqnarray*}
and
\begin{eqnarray*}
&&\mathbb{E}\left( A(T)\left[ \int_t^T\left( \int_t^TD_u\theta
(s,R_s^{t,r_0})\cdot
\theta (s,R_s^{t,r_0})ds\right) ^2du\right] ^{\frac 12}\right) \\
&\leq &C\left[\mathbb{E}\int_t^T\int_t^T\left| \frac{\partial \theta (s,R_s^{t,r_0})%
}{\partial r}\right| ^2\left| \theta (s,R_s^{t,r_0})\right| ^2\left|
D_uR_s^{t,r_0}\right| ^2dsdu\right] ^{\frac 12}<\infty ,
\end{eqnarray*}
then the third inequality holds.

\end{document}